# ThermoPix: A High-Spatial-Resolution Electronic-Photonic Temperature Sensor Array With Microsecond Row Readout

Md Rahatul Islam Udoy, *Graduate Student Member, IEEE,* Dharanidhar Dang, *Member, IEEE,* Wantong Li, *Member, IEEE*, Sumeet Kumar Gupta, *Senior Member, IEEE,* Suman Datta, *Fellow, IEEE,* and Ahmedullah Aziz, *Senior Member, IEEE*
Email: aziz@utk.edu.

*Abstract*—**This paper presents ThermoPix, a CMOS-compatible electronic–photonic architecture for high-spatial-resolution temperature sensing. The proposed system converts temperature-induced wavelength shifts in a photonic interferometric sensor into timing information that can be processed by CMOS circuitry. We use a valley photonic crystal Mach–Zehnder interferometer (VPC-MZI) as the sensing element, whose temperature-dependent spectral response is detected using an integrated waveguide photodetector and translated into a time-varying photocurrent. A CMOS readout circuit employing a phase-transition-material device performs threshold detection and generates a timing signal corresponding to the temperature-dependent crossing event. Circuit-level simulations demonstrate a temperature sensitivity of 3.15 ns/K, a row readout time of 2 µs, and a sensing power–delay product (PDP) of 0.152 fJ. The required optical power per photonic cell is 150 nW, enabling energy-efficient array operation without requiring cooling or special environmental arrangements. We also present alternative photonic layer architectures for optical power distribution across the array. In one approach, we use different tap ratios along the row, while the other uses identical tap ratios with bidirectional excitation. The resulting average photonic cell pitches are 23.26 µm and 38.52 µm, respectively. The proposed ThermoPix architecture therefore provides a scalable platform for integrated temperature sensing arrays that combine photonic sensing elements with CMOS-compatible timing-based readout.**

## I. Introduction

Temperature sensing is increasingly important in modern electronic and microsystem platforms because temperature strongly affects performance, reliability, leakage, timing, and lifetime [1], [2]. In advanced processors and heterogeneous chips, thermal hotspots can emerge locally rather than uniformly across the die, making spatially resolved thermal monitoring important for dynamic thermal management, hotspot avoidance, and reliability control [3]. Prior work on on-chip thermal monitoring has shown that sensor placement and spatial sampling strongly influence hotspot observability, while recent demonstrations of 2D thermal sensor arrays highlight the value of high-spatial-resolution mapping for resolving local CPU-core hotspots [4]. High-spatial-resolution thermal sensing is also useful beyond processors, including thermal test chips, package-level diagnostics, and distributed safety monitoring in systems such as battery packs and power electronics [5]. An array architecture is therefore desirable because a single temperature sensor can only report a local average at one point, whereas a sensor array can reconstruct temperature nonuniformity across the monitored surface [6]. This matters when the objective is not only to know the maximum temperature but also to identify where heat is concentrated and how it evolves over time [7]. Prior thermal-mapping studies explicitly note that higher sensor density improves spatial resolution and hotspot localization, although conventional dense electrical arrays can increase wiring complexity, pad count, and area overhead [8]. These considerations motivate new architectures that can scale spatial resolution without incurring a prohibitive electronic interconnect penalty.

Photonic sensing elements are attractive in this context because photonic temperature sensors can be highly sensitive, compact, and inherently immune to electromagnetic interference [9]. Reviews of photonic thermometry and silicon photonic sensing emphasize these advantages, along with compatibility with integrated photonic platforms and the ability to transduce temperature into wavelength or phase shifts with high precision [10]. However, a practical thermal imaging system cannot remain purely in the optical domain. The sensed information must ultimately be read out, processed, digitized, and interfaced with control logic, and current silicon photonics roadmaps explicitly describe photonics and electronics as complementary: photonics excels at transmission and optical transduction, while electronics provides control, readout, and digital signal processing [11], [12]. This makes an electronic-photonic co-design approach more compelling than a stand-alone photonic sensor layer. Among photonic sensing options, the valley photonic crystal Mach-Zehnder interferometer

Received nn Month 20XX; accepted nn Month 20XX. Date of publication n Month 2026; date of current version n month 2026. This work was supported in part by the ‘Funder Name’.

Md Rahatul Islam Udoy and Ahmedullah Aziz are with the Department of Electrical Engineering and Computer Science, University of Tennessee, Knoxville, TN 37996 USA (e-mail: mudoy@vols.utk.edu; aziz@utk.edu).
Dharanidhar Dang is with the Department of Computer Engineering, University of Texas at San Antonio, San Antonio, TX 78249 USA (e-mail: dharanidhar.dang@utsa.edu).
Wantong Li is with the Electrical and Computer Engineering Department, University of California, Riverside, CA 92521 USA (e-mail: wantong.li@ucr.edu).
Sumeet Kumar Gupta is with the Elmore Family School of Electrical and Computer Engineering, Purdue University, West Lafayette, IN 47907 USA (e-mail: guptask@purdue.edu).
Suman Datta is with the School of Electrical and Computer Engineering, Georgia Institute of Technology, Atlanta, GA 30332 USA (e-mail: sdatta68@gatech.edu).

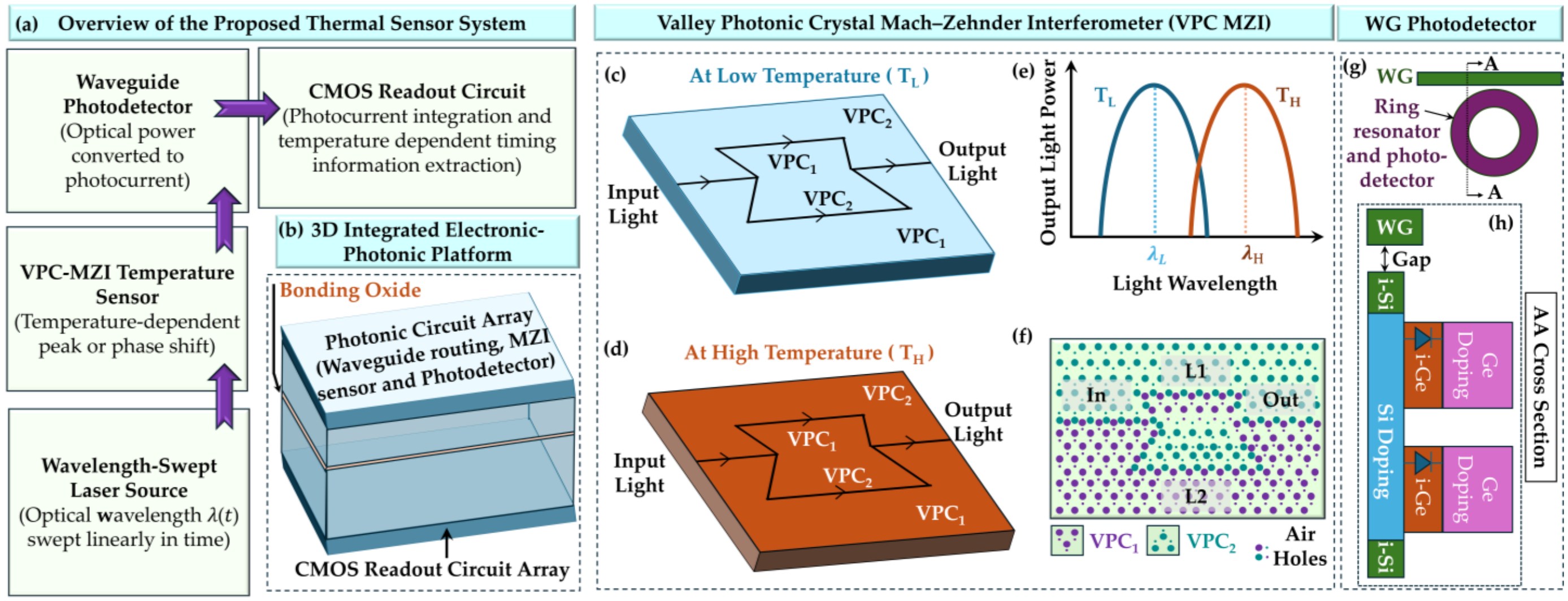

**Fig. 1:** **(a)** Simplified overview of the thermal sensor system. A wavelength-swept laser source provides input light whose wavelength is linearly swept in time and injected into the valley photonic crystal Mach-Zehnder interferometer (VPC-MZI) temperature sensor. Temperature variations cause a shift in the interference condition of the Mach-Zehnder interferometer, resulting in a change in the detected optical power. The waveguide photodetector (PD) converts the optical power into photocurrent, which is processed by the CMOS readout circuit to perform photocurrent integration and extract temperature-dependent timing information. **(b)** Conceptual illustration of the proposed three-dimensional integrated electronic-photonic platform. The upper tier contains the photonic circuit array, including waveguide routing, VPC-MZI sensors, and photodetectors. The lower tier contains the CMOS readout circuit array. The two tiers are vertically integrated through a bonding oxide layer. VPC-MZI at **(c)** low and **(d)** high temperature. **(e)** Conceptual illustration of temperature sensing through the shift of the interference peak in the output spectrum, where the peak wavelength moves from $\lambda_L$ to $\lambda_H$ as temperature increases from $T_L$ to $T_H$. **(f)** Top view layout of the VPC-MZI device formed by air holes etched in a silicon slab. The two interferometer arms (L1 and L2) are realized by zigzag-shaped waveguides formed at the boundary between $VPC_1$ and $VPC_2$. A central waveguide composed of small air holes is present but remains uncoupled due to mode mismatch [13]. **(g)** Schematic illustration of the waveguide (WG) photodetector [2]. The optical signal propagating in the WG is coupled to the ring resonator and detected by the photodetector. **(h)** A-A cross-sectional view of the WG photodetector. Optical coupling from the WG occurs across a small gap, enabling photodetection in the Ge region. The output of the VPC-MZI is coupled to a silicon strip waveguide via transverse spin matching before feeding the photodetector.

(VPC-MZI) used in this work is particularly attractive because it offers an ultracompact footprint, high forward transmittance, and compatibility with CMOS-oriented silicon photonic fabrication [13]. In addition, valley photonic crystals are widely studied as a topological photonics platform that is fabrication-friendly and promising for high-performance integrated photonic devices [13]. For an array architecture where pitch is a key design constraint, this compactness is a major advantage.

This work adopts a timing-based temperature encoding approach, where a temperature-induced wavelength shift is converted into a timing shift and detected using a mostly digital CMOS circuit instead of a fully analog amplitude readout. Time-domain sensing is attractive in scaled CMOS because it is highly digital and benefits from technology scaling without being severely affected by reduced voltage headroom, weak intrinsic gain, and device-matching limitations that challenge many analog circuits in advanced nodes [14], [15], [16]. In the proposed architecture, threshold detection is assisted by a phase-transition-material device, enabling compact and energy-efficient CMOS circuitry [17], [18]. To integrate these complementary functions effectively, the architecture employs 3D electronic–photonic integration, which allows the photonic sensing layer and CMOS readout circuitry to be optimized independently while remaining tightly coupled through dense vertical interconnects. Recent progress in heterogeneous photonic–electronic integration, including hybrid bonding and dense vertical stacking, further supports such stacked architectures for systems requiring compact photonic devices, dense sensor arrays, and local electronic readout [19].

In this paper, we propose ThermoPix, a high-spatial-resolution electronic-photonic temperature sensor array that converts temperature variations into timing information using a wavelength-swept VPC-MZI sensor, an integrated waveguide photodetector, and a CMOS timing readout circuit. The proposed architecture enables row-wise array operation with microsecond-scale readout.

The rest of the paper is organized as follows. Section II presents the system overview and background, including the VPC-MZI sensor and the waveguide photodetector. Section III describes the compact modeling framework of the VPC-MZI. Section IV presents the proposed photonic circuit layer, while Section V describes the CMOS readout circuit and array-level wiring scheme. Section VI discusses an alternative photonic layer architecture. Section VII presents PDP, pitch estimation, and benchmarking results. Finally, Section VIII concludes the paper.

## II. System Overview and Background

### A. Brief Overview

We propose ThermoPix, an architecture that converts temperature variations into timing information that can be processed by CMOS circuitry (Fig. 1(a)). We adopt a 3D electronic-photonic integration approach for this system (Fig.

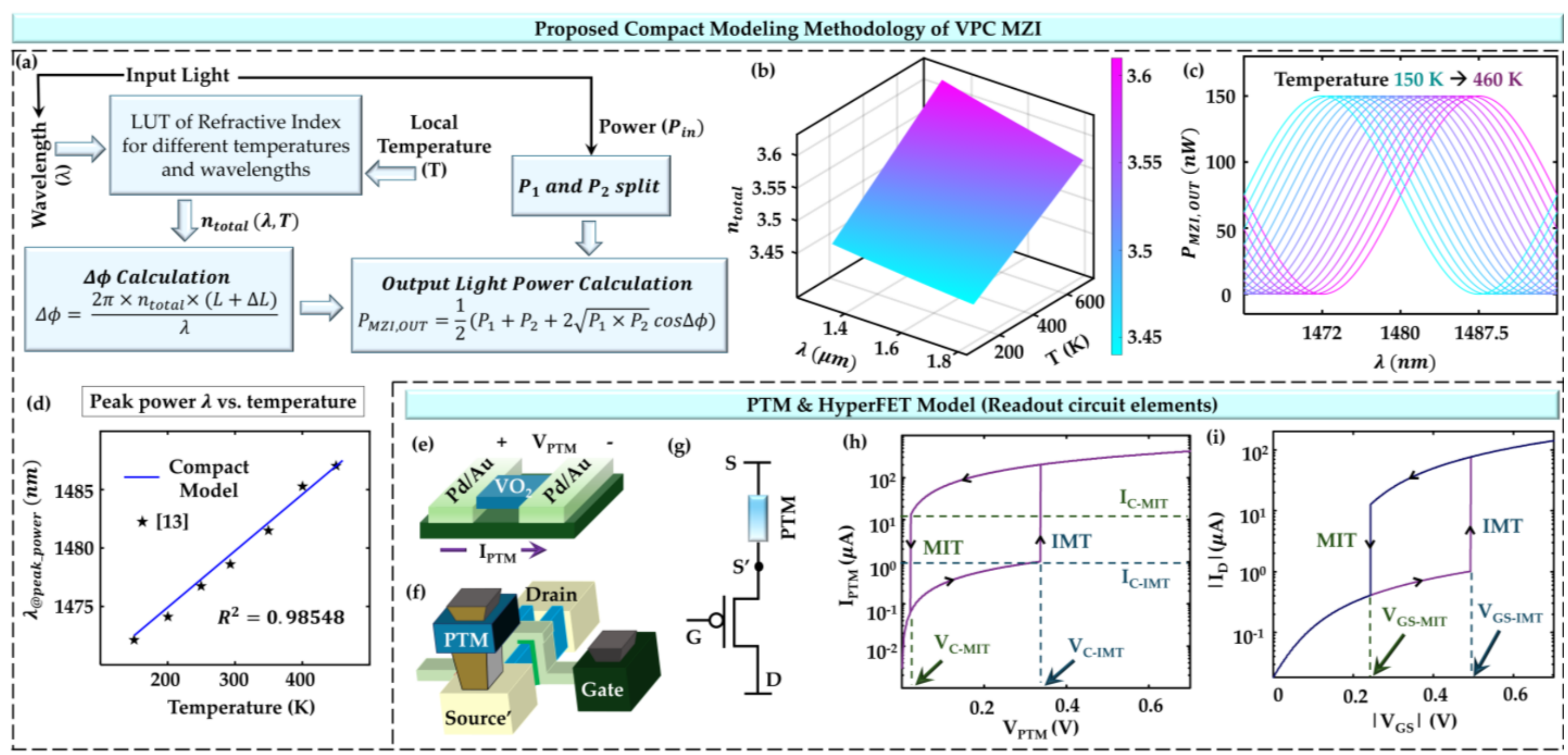

**Fig. 2: (a)** VPC-MZI compact modeling methodology showing the proposed modeling flow used to estimate the output light power of the VPC-MZI. The wavelength-swept input light and local temperature are used to obtain the refractive index from a lookup table, which is then used to calculate the phase difference and evaluate the resulting output optical power. **(b)** Refractive index data as a function of wavelength and temperature used in the model. **(c)** Simulated shift of the interference spectrum at the VPC-MZI output with temperature. **(d)** Wavelength at which the peak appears in the VPC-MZI output plotted as a function of temperature. **(e)** A $VO_2$-based phase transition material (PTM) structure. **(f)** A HyperFET structure consisting of a PTM and a MOSFET. **(g)** Circuit symbol of the HyperFET. **(h)** I-V characteristics of the PTM. When $V_{PTM}$ exceeds the critical voltage $V_{C\text{-}IMT}$, an insulator-to-metal transition (IMT) occurs. When $V_{PTM}$ drops below the critical voltage $V_{C\text{-}MIT}$, a metal-to-insulator transition (MIT) occurs. **(i)** $|I_D|$ vs $|V_{GS}|$ of the HyperFET. When $|V_{GS}|$ exceeds the critical voltage $V_{GS\text{-}IMT}$, an IMT occurs. When $|V_{GS}|$ drops below the critical voltage $V_{GS\text{-}MIT}$, an MIT occurs. Thus the device enables threshold switching.

1(b)). We begin the sensing process with a wavelength-swept laser source whose wavelength varies linearly with time. We inject the swept light into the photonic sensing structure. Temperature-dependent optical behavior in the sensor shifts the wavelength at which the output optical power peaks. During the wavelength sweep, this spectral shift appears as a shift in time. We convert the optical signal to photocurrent using an integrated waveguide photodetector. We then process this photocurrent using a CMOS readout circuit to extract temperature-dependent timing information. The photonic sensing elements and the CMOS readout circuits are arranged on separate tiers of a 3D integrated platform. This type of multilayer integration has been experimentally demonstrated in the literature [20]. The photonic circuit array resides on the upper tier, while the CMOS readout circuit array resides on the lower tier. The two tiers are connected through a bonding oxide layer. This architecture enables compact integration of the sensing and readout functions for high spatial resolution temperature monitoring.

### *B. VPC-MZI and WG Photodetector*

We perform temperature sensing using a valley photonic crystal Mach-Zehnder interferometer (VPC-MZI) shown in Fig. 1(c)-(f). The VPC-MZI device used in this work is adapted from [13]. This device is suitable for the proposed ThermoPix system because its compact footprint and CMOS-compatible silicon photonic platform allow dense on-chip integration of large sensor arrays. The device is implemented on a silicon slab of thickness 0.22 µm patterned with a honeycomb lattice of air holes (Fig. 1 (f)). The photonic crystal lattice constant is a=0.44 µm. Each unit cell contains two air holes. By breaking the spatial inversion symmetry through unequal hole sizes, a photonic bandgap is opened. Two valley photonic crystal domains, $VPC_1$ and $VPC_2$, are defined by opposite arrangements of the small-radius and large-radius holes. In $VPC_1$, the large-radius hole is 0.12 µm and the small-radius hole is 0.04 µm. In $VPC_2$, these two radii are interchanged. Because these two domains possess opposite valley properties, a topological edge state is formed at the boundary between them. Waveguides are created by placing $VPC_1$ and $VPC_2$ adjacent to each other. The VPC-MZI uses zigzag-shaped boundaries. The interferometer consists of an input topological waveguide followed by two arms labeled $L_1$ and $L_2$. The input waveguide length is L = 8a, while the two arms have lengths $L_1$ = 18a and $L_2$ = 23a. Therefore, the path length difference is ΔL = 5a. A central waveguide composed of small air holes is also present, but it remains uncoupled because of mode mismatch. The footprint of this device is 9.26 $\mu$m × 7.99 $\mu$m..

Temperature sensing arises from the thermo-optic effect of silicon. As temperature increases, the refractive index of silicon increases. This changes the phase accumulated along the interferometer arms and shifts the interference peak in the output spectrum toward longer wavelength as shown in Fig. 1 (c)-(e).

The output optical signal is detected using an integrated waveguide photodetector (WG-PD) whose structure is adapted from [21]. The compact footprint, high responsivity (~1 A/W), and CMOS-compatible integration of this detector make it well suited for the proposed ThermoPix architecture. The topological waveguide mode supported at the interface of the VPC domains can be efficiently coupled to a conventional silicon strip waveguide using the transverse spin matching (TSM) mechanism reported in [22]. This enables seamless routing of the optical signal from the VPC-MZI to the downstream photodetection circuitry implemented on standard silicon photonic waveguides. The detector consists of a germanium-on-silicon resonant photodetector evanescently coupled to a silicon bus waveguide as illustrated in Fig. 1(g)-(h). The bus waveguide width is 400 nm, which ensures single-mode operation. Optical power propagating in the silicon waveguide is coupled into the detector through evanescent coupling across a small gap between the waveguide and the resonant cavity. The detector employs a germanium absorption region integrated on silicon, where silicon doping, intrinsic germanium (i-Ge), and doped germanium form a p–i–n junction structure. The photodetector uses a ring resonator cavity with a radius of 4.5 µm, enabling a compact footprint and enhanced optical absorption. The separation between the germanium and silicon outer radius is 1.5 µm, which allows a controlled overlap between the optical mode and the germanium absorption region. When optical power is absorbed in the germanium region, electron-hole pairs are generated, producing photocurrent. This photocurrent is subsequently processed by the CMOS readout circuit to extract temperature-dependent timing information. Because the wavelength-swept sensing mechanism converts temperature-dependent spectral shifts into time-varying optical power, the photodetector directly produces a temperature-dependent photocurrent waveform suitable for CMOS timing extraction.

## III. VPC-MZI Compact Modeling

To enable circuit-level simulation of the proposed thermal sensing system, we develop a Verilog-A based compact model of the valley photonic crystal Mach-Zehnder interferometer (VPC-MZI). The objective of this model is to capture the temperature-dependent optical response of the interferometer while maintaining computational efficiency and compatibility with electronic circuit simulation. The overall modeling framework is illustrated in Fig. 2(a). In the compact model, the optical excitation is represented by a wavelength-swept input $\lambda(t)$ with input optical power $P_{in}$. These quantities serve as model inputs to the interferometer. The input optical power is distributed between the two interferometer arms such that $P_1 = P_2 = P_{in}/2$. The temperature sensing mechanism of the VPC-MZI originates from the thermo-optic effect of silicon. As temperature varies, the refractive index of silicon changes, which alters the optical phase accumulated along the interferometer arms. To efficiently capture this behavior, we construct a lookup table (LUT) to represent the effective refractive index of silicon as a function of wavelength and temperature. The LUT stores values of the effective refractive index $n_{total}(\lambda, T)$ using reported optical material data [13], [23]. Using a lookup table avoids repeated electromagnetic simulations during circuit-level analysis and enables efficient evaluation of the interferometer response. The refractive index determines the optical phase accumulated along the interferometer arms. Because the two arms have a path length difference $\Delta L$, a phase difference is generated between the two optical signals. The phase difference is calculated as

$$\Delta\phi = \frac{2\pi}{\lambda} \times n_{total} \times \Delta L \tag{1}$$

The output optical power of the interferometer is then obtained using the standard interference relation of a Mach-Zehnder interferometer

$$P_{MZI,\,OUT} = \frac{1}{2}\left(P_1 + P_2 + 2\sqrt{P_1 \times P_2}\,cos\Delta\phi\right) \tag{2}$$

This formulation allows the compact model to compute the temperature-dependent optical output of the interferometer for each wavelength of the input sweep. Fig. 2(b) shows the refractive-index data used in the model, illustrating the dependence of the silicon refractive index on wavelength and temperature. As temperature increases, the refractive index increases due to the thermo-optic effect. This variation modifies the phase difference between the interferometer arms and consequently shifts the interference spectrum. The resulting spectral response of the VPC-MZI predicted by the compact model is shown in Fig. 2(c). As temperature increases, the interference peak shifts toward longer wavelengths. For sensing operation, the key parameter is the wavelength corresponding to the maximum output optical power. Therefore, the wavelength at peak power (WPP) of the interference spectrum is extracted for each temperature value. The relationship between WPP and temperature obtained from the compact model is shown in Fig. 2(d). The model results closely follow the reported device characteristics, with a coefficient of determination $R^2 = 0.9855$. The optical output power of the VPC-MZI is applied to the waveguide photodetector model, where the optical signal is converted into photocurrent. This photocurrent is then processed by the CMOS readout circuit. Because the VPC-MZI model relies on analytical relations and precomputed lookup tables, it is inherently computationally efficient by construction, making it suitable for circuit-level simulation of large sensor arrays [24].

## IV. Photonic Circuit Layer

The proposed photonic circuit layer distributes the wavelength-swept optical signal to the array of sensing elements and enables scalable operation of the ThermoPix architecture. As shown in Fig. 3(a), the layer consists of an $m \times n$ array of photonic cells interconnected through a network of waveguides and directional tap couplers. The optical excitation is provided by a wavelength-swept laser source whose wavelength varies linearly with time, as illustrated in Fig. 4(a). In this work, the laser wavelength sweeps from approximately 1470 nm to 1490 nm over a 2 $\mu s$ interval, corresponding to a sweep rate of approximately 15 $nm/\mu s$. Such high-speed wavelength sweeping is experimentally feasible. Recent demonstrations of wavelength-swept DFB laser arrays have achieved continuous (gap-free) tuning ranges up to 60 nm with sweeping speeds exceeding 24 $nm/\mu s$, confirming the practicality of microsecond-scale wavelength sweeps for optical sensing systems [25].

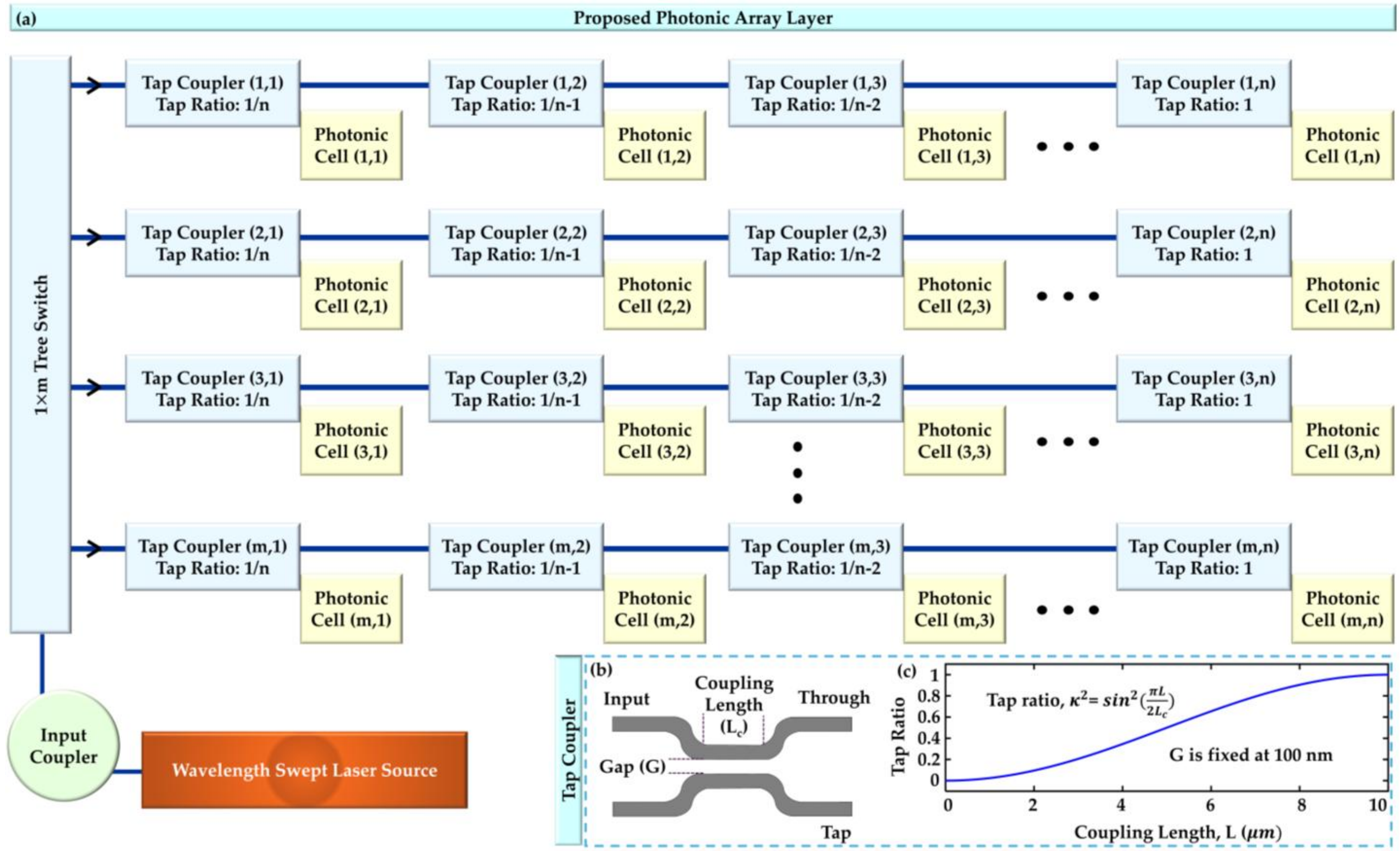

**Fig. 3: (a)** Photonic circuit layer consisting of an m × n array of photonic cells connected through a series of tap couplers. A wavelength-swept laser source feeds the circuit through an input coupler and a 1 × m tree switch. The tree switch routes the full optical power from the laser to one row at a time, similar to a rolling-shutter operation. The tap ratios are chosen so that each photonic cell receives approximately equal optical power. Each tap coupler extracts a fraction of the optical power and routes the remaining power to the next stage. Each photonic cell consists of a VPC-MZI sensor, a WG photodetector, and a through-oxide-via (TOV) landing pad. The TOV landing pad provides the vertical electrical connection to the CMOS circuit layer. **(b)** Directional tap coupler structure. The Input port receives the incoming optical signal. The Through port carries the remaining optical power after coupling. The Tap port extracts a fraction of the optical power determined by the coupling length and the gap. **(c)** Different tap ratios are achieved by varying the coupling length of the tap coupler.

The optical signal is injected into the photonic circuit layer through an input coupler and then routed to the array using a $1 \times m$ tree switch. Here, the notation $1 \times m$ denotes an optical switching structure with one input port and $m$ output ports. Silicon photonic switching networks capable of routing a single optical input to multiple output ports have been experimentally demonstrated [26], making them suitable for implementing the row-selective routing required in the proposed array. The tree switch directs the optical power from the laser to one row of the sensor array at a time, enabling sequential row operation similar to a rolling-shutter mechanism. This approach allows the entire array to share a single optical source while maintaining scalable addressing of the sensing elements. Because only one row is illuminated at a time, the required optical power scales with the number of cells in a row rather than the total number of cells in the array. In this work, each photonic cell requires approximately 150 nW of optical power. Therefore, the required laser power for a row is approximately $n \times 150$ nW plus distribution losses in the photonic routing network. We determine the required optical power for each cell based on the wavelength sweep duration, the capacitance of the photodetector, and the photodetector dark current. We present the detailed derivation of this power requirement in the next section.

Within each row, optical power is distributed using a sequence of directional tap couplers. Each tap coupler extracts a fraction of the propagating optical power and routes it to a corresponding photonic cell while allowing the remaining power to continue toward the next stage. The tap ratios are selected so that approximately equal optical power is delivered to every photonic cell along the row. If the optical power entering a row is $P_{row,in}$ and the row contains $n$ photonic cells, the first tap coupler extracts a fraction $1/n$ of the optical power and delivers $P_{row,in}/n$ to the first cell. The remaining optical power $P_{row,in} \times \frac{n-1}{n}$ propagates to the next coupler. The second coupler extracts a fraction $1/(n-1)$ of the remaining power, which again yields approximately $P_{row,in}/n$ delivered to the second cell. This process continues along the row such that the $k$-th tap coupler extracts a fraction $1/(n-k+1)$ of the remaining optical power. Consequently, each photonic cell receives approximately the same optical power despite the progressive extraction along the waveguide.

The tap couplers used in this work are directional couplers, illustrated in Fig. 3(b). Optical coupling occurs over a coupling region of length $L_c$ between two parallel waveguides separated by a gap $G$. The fraction of optical power transferred to the tap port depends on the coupling length and the waveguide separation. For a fixed waveguide gap, the tap ratio $\kappa^2$ follows

$$\kappa^2 = sin^2\left(\frac{\pi L}{2L_c}\right) \tag{3}$$

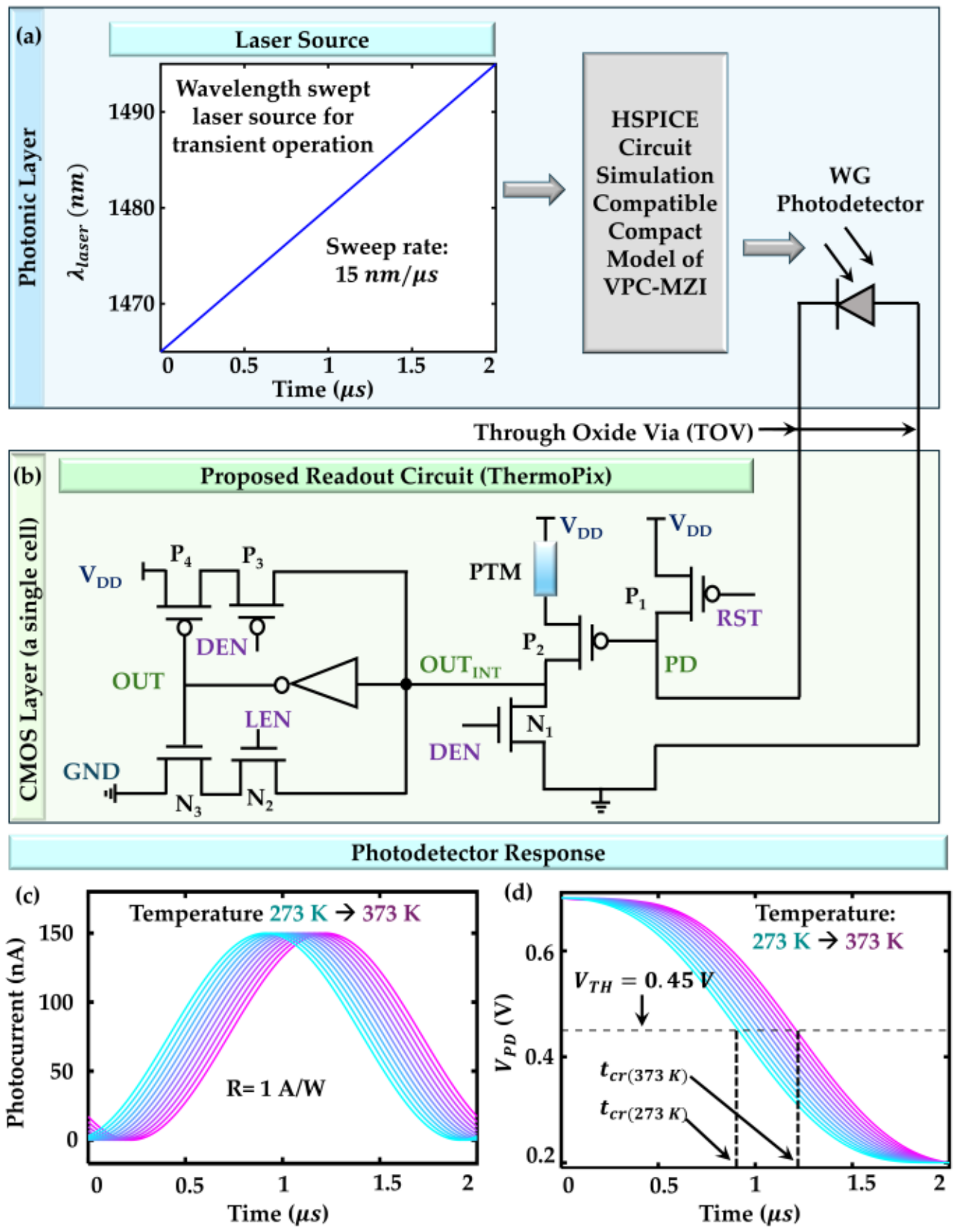

**Fig. 4: (a)** Photonic layer transient operation. A wavelength-swept laser performs a linear sweep in time and drives the VPC-MZI sensor. This transient sweep converts the temperature-dependent spectral shift of the VPC-MZI into a time-domain signal. The proposed compact model of the VPC-MZI enables HSPICE-compatible circuit simulation. The optical signal is detected by a WG photodetector and the resulting electrical signal is transferred to the CMOS layer through TOV. **(b)** Proposed readout circuit for a single photonic cell. Transistor $P_1$ is the reset transistor, Transistor $P_2$ and PTM forms a HyperFET, Transistor $N_1$ is used to discharge the $OUT_{INT.}$ $N_2$, $N_3$, $P_3$, $P_4$ and the inverter forms a latch. RST, LEN, DEN are pulse signals provided. RST, LEN, and DEN are externally applied pulse signals. The detailed operation of the circuit is shown in Fig. 5. **(c)** Photocurrent generated by the WG photodetector during the wavelength sweep. The waveform shifts in time with temperature due to the temperature-dependent spectral shift of the VPC-MZI output. **(d)** Voltage evolution at the sensing node. The threshold crossing time changes with temperature and is used to extract the temperature information. The temperature range is reduced compared to the full sensing range of the VPC-MZI (Fig. 2(c)) because this range matches the target application of on-chip monitoring and the operating limits of the CMOS readout circuit.

Here $L$ is the design parameter representing the interaction length of the directional coupler, while $L_c$ is the characteristic coupling length determined by the waveguide geometry and gap. By varying the coupling length while keeping the waveguide geometry fixed, different tap ratios can be realized. Fig. 3(c) illustrates the variation of the tap ratio as a function of coupling length for a fixed gap of $G$ = 100 nm. A coupling length 10 $\mu m$ gives a full tap ratio. Each photonic cell integrates a VPC-MZI temperature sensor and a WG-PD. There is a through-oxide-via (TOV) landing pad in each cell, which provides the vertical electrical connection to the CMOS readout circuitry located in the lower tier of the 3D integrated platform.

## V. CMOS Circuit Layer

### A. Circuit Configuration

The proposed ThermoPix readout circuit is shown in Fig. 4(b). The circuit receives the photocurrent generated by the waveguide photodetector. The design integrates conventional CMOS devices with a phase-transition material (PTM) based switching element to enable threshold-triggered timing detection. Phase-transition materials exhibit abrupt changes in resistivity and can be triggered electrically. A representative PTM structure is shown in Fig. 2(e). As illustrated in Fig. 2(h), a PTM remains in an insulating state with high resistance at low bias. When the applied voltage exceeds a critical value $V_{C-IMT}$, an insulator-to-metal transition (IMT) occurs at a corresponding current $I_{C-IMT}$. In the insulating state, the resistivity of the material is several orders of magnitude larger than the metallic resistivity. When the voltage is reduced below the critical value $V_{C-MIT}$, the device undergoes a metal-to-insulator transition (MIT) at current $I_{C-MIT}$. A variety of PTM materials exhibiting different transition voltages and hysteresis behaviors have been reported in the literature [27], [28], [29]. When a PTM device is placed in series with the source terminal of a conventional MOSFET, the resulting device is referred to as a HyperFET, as illustrated in Fig. 2(f)-(g). We use PMOS as the FET. The HyperFET combines standard FET conduction with the abrupt switching behavior of the PTM. For small $|V_{GS}|$, the MOSFET remains off and the PTM stays in its high-resistance state, effectively introducing a large source resistance. When $|V_{GS}|$ exceeds a critical value $V_{GS-IMT}$, the PTM undergoes the IMT and the source resistance abruptly decreases. As $|V_{GS}|$ is reduced below $V_{GS-MIT}$, the PTM returns to its high-resistance state through the metal-to-insulator transition. Numerous circuit techniques utilizing HyperFET devices have been demonstrated in prior work. In the proposed ThermoPix circuit, we exploit this abrupt switching behavior to implement a threshold-triggered timing detector. The PTM-HyperFET element acts as a sharp switching device that converts the gradual voltage change produced by photodetector integration into a well-defined digital transition.

In the circuit, transistor $P_1$ acts as a reset device that initializes the photodetector node before each sensing cycle. Transistor $P_2$ together with the PTM forms the HyperFET switching element described earlier. Transistor $N_1$ is used to discharge the intermediate node $OUT_{INT}$ during the sensing sequence. The remaining transistors $N_2$, $N_3$, $P_3$, and $P_4$ together with the inverter form a latch. Three external control signals are used to operate the circuit. The reset signal RST initializes the circuit at the beginning of the sensing cycle, while the enable signals LEN and DEN control the integration and detection phases of operation.

### B. Photodetector Response

The photocurrent generated by the waveguide photodetector during the wavelength sweep is shown in Fig. 4(c). Although the VPC-MZI sensor adapted in this work can support a much wider temperature range, we restrict the operating range to 273 K-373 K. This range is sufficient for the target applications of the proposed ThermoPix architecture, such as high-spatial-resolution on-chip temperature monitoring, where temperature variations typically occur within a limited range. In addition, other photonic and electronic components in the system are not designed to operate reliably at the extremely high temperatures supported by the VPC-MZI structure. Therefore, limiting the sensing range to 273 K-373 K provides a practical operating window that matches the requirements of the intended application while remaining compatible with the photonic and CMOS circuitry. As the wavelength of the input laser varies linearly with time, the optical output of the VPC-MZI produces a time-varying optical power at the detector. Because the interference peak of the VPC-MZI shifts with temperature, the resulting photocurrent waveform also shifts in time, as illustrated in Fig. 4(c) for different temperatures. Assuming equal optical power in the two arms of the interferometer, the optical power incident on the photodetector can be expressed as

$$P_{PD,in}(t,T) = \frac{1}{2}\left(P_1 + P_2 + 2\sqrt{P_1 P_2}\cos\left(\omega t + \phi(T)\right)\right)$$

where $P_1$ and $P_2$ are the optical powers from the two interferometer arms, $\omega$ represents the angular sweep rate, and $\phi(T)$ is the temperature-dependent phase shift introduced by the VPC-MZI. The waveguide photodetector converts this optical signal into electrical current according to

$$I_{PD}(t,T) = I_{dark} + RP_{PD,in}(t,T)$$

where $R$ is the photodetector responsivity and $I_{dark}$ is the dark current. Using $R = 1\,A/W$ and $P_1 = P_2 = 75\,nW$, the photodetector current becomes

$$I_{PD}(t,T) = I_{dark} + 75\text{nA}\left[1 + \cos\left(\omega t + \phi(T)\right)\right]$$

Thus the photocurrent varies between $I_{dark}$ and $I_{dark} + 150\,nA$ during the wavelength sweep. The photodetector current is integrated on the sensing node capacitance to generate the voltage waveform shown in Fig. 4(d). The voltage evolution follows

$$\frac{dV_{PD}(t,T)}{dt} = -\frac{I_{PD}(t,T)}{C_{PD}}$$

where $C_{PD}$ is the capacitance of the sensing node. Integrating the above expression gives

$$V_{PD}(t,T) = V_{RST} - \frac{1}{C_{PD}}\int_0^t I_{PD}(\tau,T)\,d\tau$$

Substituting the photocurrent expression yields

$$V_{PD}(t,T) = V_{RST} - \frac{I_{dark} + 75\text{ nA}}{C_{PD}}t - \frac{75\text{ nA}}{\omega C_{PD}}\left[\sin\left(\omega t + \phi(T)\right) - \sin\left(\phi(T)\right)\right]$$

This equation describes the voltage trajectory at node $V_{PD}$ during the wavelength sweep. As the optical waveform shifts in time with temperature, the voltage waveform in Fig. 4(d) also shifts accordingly. The sensing circuit detects the moment when the integrated voltage crosses the threshold voltage of the HyperFET. The threshold-crossing time $t_{cr}$ is therefore obtained from

$$V_{PD}(t_{cr},T) = V_{TH}$$

The analytical relation-ship between phase shift and threshold-crossing time is shown in supplementary material S1. Because $\phi(T)$ varies with temperature, the crossing time $t_{cr}$ also shifts in time, enabling temperature sensing through timing detection. To verify that the photocurrent is sufficient to produce the required voltage swing within the sweep duration, we estimate the integration time using realistic parameters. The sensing node capacitance is approximately $C_{PD} = 600\,fF$ (which includes photodetector capacitance, transistor $P_2$ gate capacitance, TOV capacitance [20], routing parasitics). The required voltage swing is $\Delta V = V_{DD} - V_{TH} = 0.7 - 0.45 = 0.25\,V$. The nominal photocurrent generated by the detector is approximately 150 $nA$. The minimum dark current reported for the adapted detector is 2.03 $nA$, while the dark current can increase by approximately 34 × at $100^0$ C compared to $100^0$ C [30], giving an upper bound of approximately 68 $nA$. Therefore the total integration current lies in the range

$$I_{total} = I_{photo} + I_{dark}$$

$$152.03\text{ nA} \leq I_{total} \leq 218\text{ nA}$$

The integration time required to generate the voltage swing becomes

$$t = \frac{C_{PD}\Delta V}{I_{total}}$$

which yields

$$0.69\ \mu\text{s} \leq t \leq 1\ \mu\text{s}$$

This integration time is well within the wavelength sweep duration (2 $\mu s$), ensuring that the sensing node voltage can reach the threshold within the transient sweep interval.

Although dark current contributes to the integrated current and can shift the threshold-crossing time, its effect can be mitigated using correlated double sampling in the readout circuitry. Correlated double sampling suppresses offsets associated with dark current and low-frequency noise, enabling more accurate extraction of the temperature-dependent timing information.

The photocurrent level also determines the required optical power incident on the photodetector. For a responsivity of $R = 1\text{ A/W}$ a nominal photocurrent of approximately $I_{photo} \approx 150\text{ nA}$ corresponds to an optical power $P_{PD,in} \approx 150\text{ nW}$ per photonic cell. Since the optical power is distributed along a row of $n$ photonic cells, the required row input power becomes $P_{row,in} = n \times P_{PD,in}$. As an example, for a row containing 64 cells, $P_{row,in} = 64 \times 150\text{ nW} \approx 9.6\ \mu\text{W}$. Allowing additional margin for waveguide and coupler losses, a laser power of approximately $P_{laser} \approx 15\ \mu\text{W}$ provides sufficient optical power for reliable operation. Here allowable loss is 36%.

### *C. Circuit Operation and Output Waveforms*

We evaluate the operation of the proposed ThermoPix circuit using HSPICE, an industry-grade circuit simulation tool. The CMOS transistors are modeled using the Predictive Technology Model (PTM) [31]. To represent the phase-transition material (PTM), we incorporate the behavioral circuit model of vanadium dioxide ($VO_2$) reported in [27]. The PTM model captures the abrupt insulator-to-metal transition (IMT), metal-to-insulator transition (MIT), and the associated hysteresis behavior. The PTM simulation parameters used in this work are summarized in Table I. The simulated output waveforms illustrating the circuit operation are shown in Fig. 5(a)-(t) for three representative temperatures: 273 K, 327 K, and 373 K. The operation of the ThermoPix circuit begins with the reset phase. During this phase, the reset signal activates transistor $P_1$, charging the photodetector node PD to $V_{DD}$. Because $P_2$ is a PMOS device whose gate is tied to the PD node, the high gate voltage keeps $P_2$ in the off state, resulting in a large channel resistance. Consequently, in the series combination of $P_2$ and the PTM, most of the voltage drop occurs across $P_2$ according to the voltage divider rule. The PTM therefore does not receive sufficient voltage to trigger the insulator-to-metal transition (IMT) and remains in the insulating state. After reset, the circuit enters the integration phase. The reset transistor $P_1$ is turned off, and $V_{PD}$ begins to decrease as photocurrent is generated by the waveguide photodetector (Fig. 5(a)). The temporal evolution of $V_{PD}$ follows the integration relation described in the previous section. Because the photocurrent waveform shifts in time with temperature, the rate and timing of the voltage drop depend on the temperature of the VPC-MZI sensor.

We design the ThermoPix architecture to measure the time at which the sensing node voltage crosses the HyperFET threshold. Since the integration process driven by the photocurrent eventually causes the threshold to be reached, the sensing task is to determine the exact timing of this event. To detect this moment, the control signals DEN and LEN are applied as periodic pulses during the sensing interval as shown in Fig. 5 (b)-(c), (m)-(n). These pulsed control signals periodically enable the evaluation path of the circuit, allowing the circuit to check whether the threshold condition has been reached. The DEN signal enables the detection path, allowing the circuit to evaluate the state of the HyperFET and the LEN signal activates the latch network. Once the threshold crossing occurs, the circuit produces the corresponding output transition that marks the sensing event. Initially, the pulsing of DEN and LEN begins after a predefined delay from the start of the wavelength sweep. This delay is selected based on the expected earliest threshold-crossing time corresponding to the lower bound of the operating temperature range (273 K). When the HyperFET threshold has not yet been reached, the PTM remains in the insulating state and the resistance of the $P_2$-PTM path remains high. Under this condition, the intermediate node $OUT_{INT}$ stays in the low state and the final output remains unchanged (Fig. 5(d)-(i), (o)-(p)). As the integration proceeds, the decreasing $V_{PD}$ eventually reaches a level where the magnitude of the HyperFET gate-source voltage $|V_{GS}|$ exceeds the critical transition voltage $|V_{GS-IMT}|$. At this moment, the PTM undergoes the insulator-to-metal transition, drastically reducing its resistance (Fig. 5 (j)-(l),(q)). The upper branch of the circuit now provides a low-resistance path to charge the $OUT_{INT}$ node. Consequently, this node transitions from low to high, which causes the inverter output OUT to fall from high to low (Fig. 5 (g)-(i)). Once this transition occurs, subsequent pulses of DEN and LEN produce a periodic output waveform. However, only the first transition corresponds to the actual sensing event.

**TABLE I: PTM simulation parameters**

| Parameters | Definition | Value |
|---|---|---|
| $L_{PTM}, W_{PTM}$ | Length, Width | 45 nm, 55 nm |
| $T_{PTM}$ | Thickness | 20 nm |
| $A_{PTM}$ | Cross-sectional area | 1100 nm$^2$ |
| $J_{C\text{-}IMT}$ | Critical current density for IMT | $8\times10^4$ A/cm$^2$ |
| $J_{C\text{-}MIT}$ | Critical current density for MIT | $1\times10^6$ A/cm$^2$ |
| $\rho_M$ | Resistivity in metallic state | $5\times10^{-3}$ $\Omega$.cm |
| $\rho_{INS}$ | Resistivity in insulating state | 1 $\Omega$.cm |

The timing resolution of the system depends on the pulsing frequency of the DEN and LEN control signals. A higher pulsing rate enables finer temporal quantization of the threshold-crossing event, resulting in improved temperature resolution.

From Fig. 5(d)-(i), it can be observed that the transitions of $OUT_{INT}$ and OUT occur first for 273 K, followed by 327 K, and finally for 373 K. This behavior arises because the threshold-crossing time is lowest for 273 K and highest for 373 K within the operating range considered, as theoretically established in the previous section. The three temperatures shown in the waveform plots are presented only as representative examples for clarity. In the actual analysis, the temperature is swept from 273 K to 373 K, and the corresponding output-flipping time is recorded. The extracted threshold-crossing time as a function of temperature is plotted in Fig. 5(u).

The threshold voltage used for sensing is determined by the PTM characteristics listed in Table I. Because the switching behavior of the HyperFET depends on the PTM transition parameters, the effective threshold of the circuit can be tuned through the PTM properties. In particular, the threshold level can be adjusted by modifying the physical dimensions of the PTM device, which changes its effective resistance and transition voltage. In addition, PTM materials with different critical current densities and resistivities have been reported in the literature [32], and these variations can also be utilized to tune the threshold level of the sensing circuit.

### *D. Array Level Wiring Diagram*

The array-level wiring of the CMOS readout layer is illustrated in Fig. 5(v). The architecture employs a row-wise activation scheme that is synchronized with the operation of the photonic circuit array, where one row of thermopixels is addressed at a time. In this configuration, the control signals RST, DEN, and LEN are distributed horizontally along each row of the ThermoPix array. These signals are shared by all thermopixels within the same row and are used to sequentially activate one row of pixels at a time. When a particular row is selected, the corresponding thermopixels perform the sensing operation simultaneously while the other rows remain inactive. This row-wise addressing approach simplifies the control circuitry and allows the array to scale to a large number of sensing elements [33], [34]. Each thermopixel produces an output signal OUT that indicates the occurrence of the threshold-crossing event in the sensing circuit. These OUT signals are routed vertically along the columns of the array. All pixels within the same column share a common column line, which carries the OUT signal toward the column-end circuitry. This column-wise signal routing enables centralized processing of the timing information generated by the pixels while minimizing routing complexity within the array. At the bottom of each column, a column-end circuit processes the output signals from the corresponding column. The column-end circuitry contains a digital counter together with simple control logic used to measure the threshold-crossing time generated by the thermopixel. The counting operation begins when the wavelength sweep of the laser source starts. This start signal is distributed to all column counters and acts as the reference time for the sensing operation. As the sensing process proceeds, the integrated voltage within the thermopixel eventually reaches the switching threshold, which causes the output signal OUT to transition. The first falling edge of the OUT signal is used as a stop signal for the counter. At this moment, the counter value represents the elapsed time between the start of the wavelength sweep and the threshold-crossing event, which corresponds to the sensed temperature. After the switching event, the output signal begins to oscillate due to the pulsed DEN and LEN control signals applied during the sensing phase. However, only the first transition is used for temperature extraction. The subsequent output pulses do not restart or affect the counter because the counting operation can only be initiated by the global start signal. This mechanism ensures robust extraction of the threshold-crossing time while preventing multiple output transitions from corrupting the measurement.

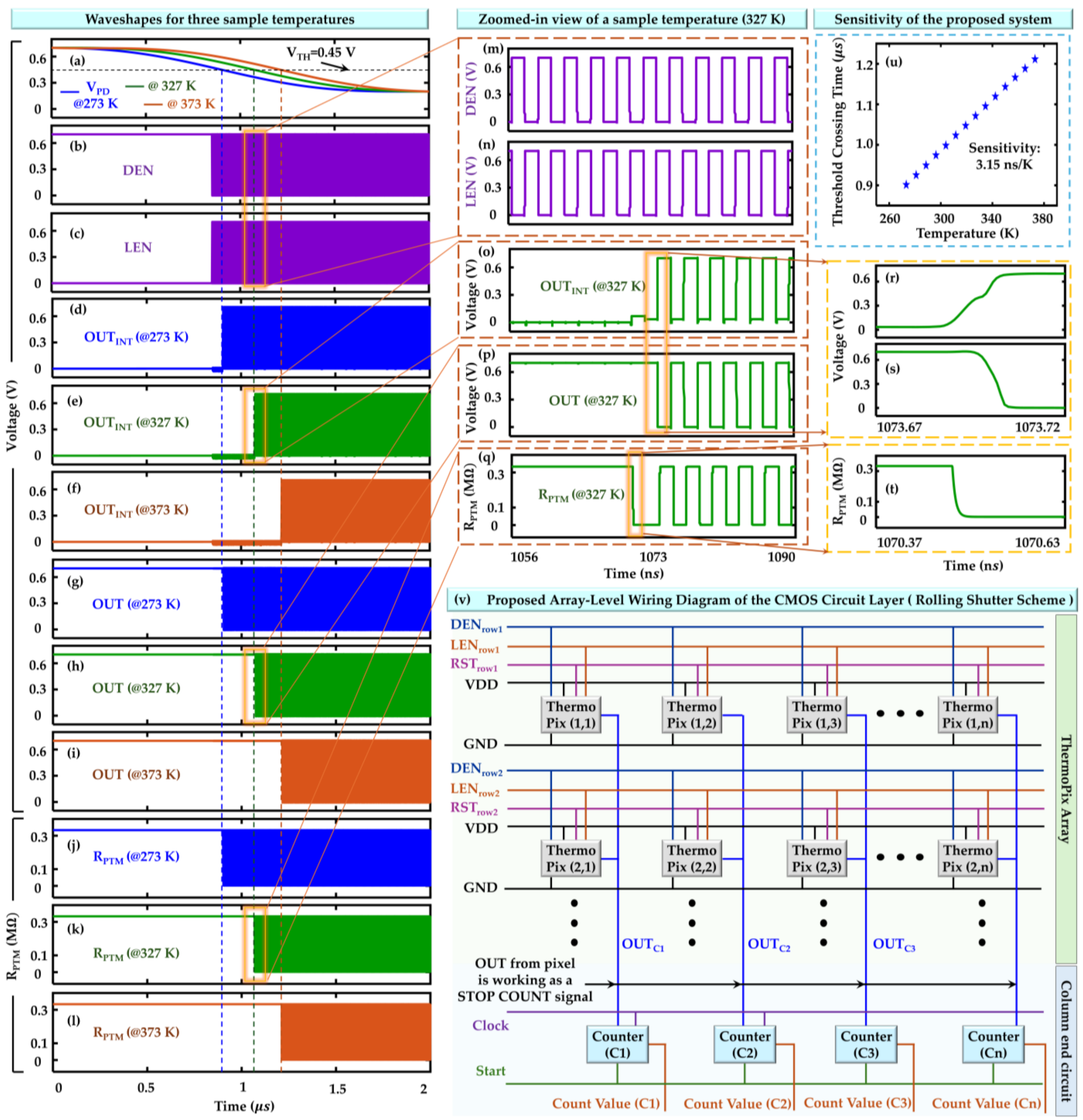

**Fig. 5: (a)-(l)** Simulated waveforms of the ThermoPix circuit shown in Fig. 4(b) at three sample temperatures (273 K, 327 K, and 373 K). The waveforms are shown after the reset stage. During reset, transistor $P_1$ is turned on and sets $V_{PD}$ to $V_{DD}$. $P_1$ is then turned off and $V_{PD}$ begins to decrease due to the photocurrent of the photodetector. The voltage $V_{PD}$ eventually crosses the threshold voltage $V_{TH}$ set by the HyperFET. Control signals DEN and LEN operate in a pulsed sensing-latching sequence. When DEN=High and LEN=Low, the nodes $OUT_{INT}$ remain Low and OUT remains High. If $V_{PD}$ crosses $V_{TH}$ before the sensing-latching transition (DEN=Low and LEN=High), the PTM switches; Both $OUT_{INT}$ and OUT flip and start pulsing. If $V_{PD}$ does not cross $V_{TH}$ before this transition, the PTM does not switch and the output remains unchanged. Because the crossing time is unknown, the DEN and LEN pulses are repeated until the laser completes its wavelength sweep. Panels **(d), (g), and (j)** show early switching at 273 K. Panels **(e), (h), and (k)** show later switching at 327 K. Panels **(f), (i), and (l)** show the latest switching at 373 K**. (m)-(n)** Zoomed-in view of the DEN and LEN control pulses. **(o)-(t)** Zoomed-in waveforms for a sample temperature of 327 K showing the internal node voltages and PTM resistance transition during switching. **(u)** Extracted threshold-crossing time as a function of temperature, showing the sensitivity of the proposed system. **(v)** Proposed Array-level wiring of the CMOS circuit layer using a rolling-shutter scheme. The control signals RST, DEN, and LEN are shared along each row and activate one row of thermopixels at a time. The OUT signal from each thermopixel is shared along the column. The column-end circuit contains a counter and control logic. A start signal begins counting when the wavelength sweep starts. The first falling edge of OUT stops the counter when switching occurs. After switching, OUT begins pulsing, but only the first transition is used for counting. Subsequent transitions do not restart the counter because counting can only be initiated by the start signal.

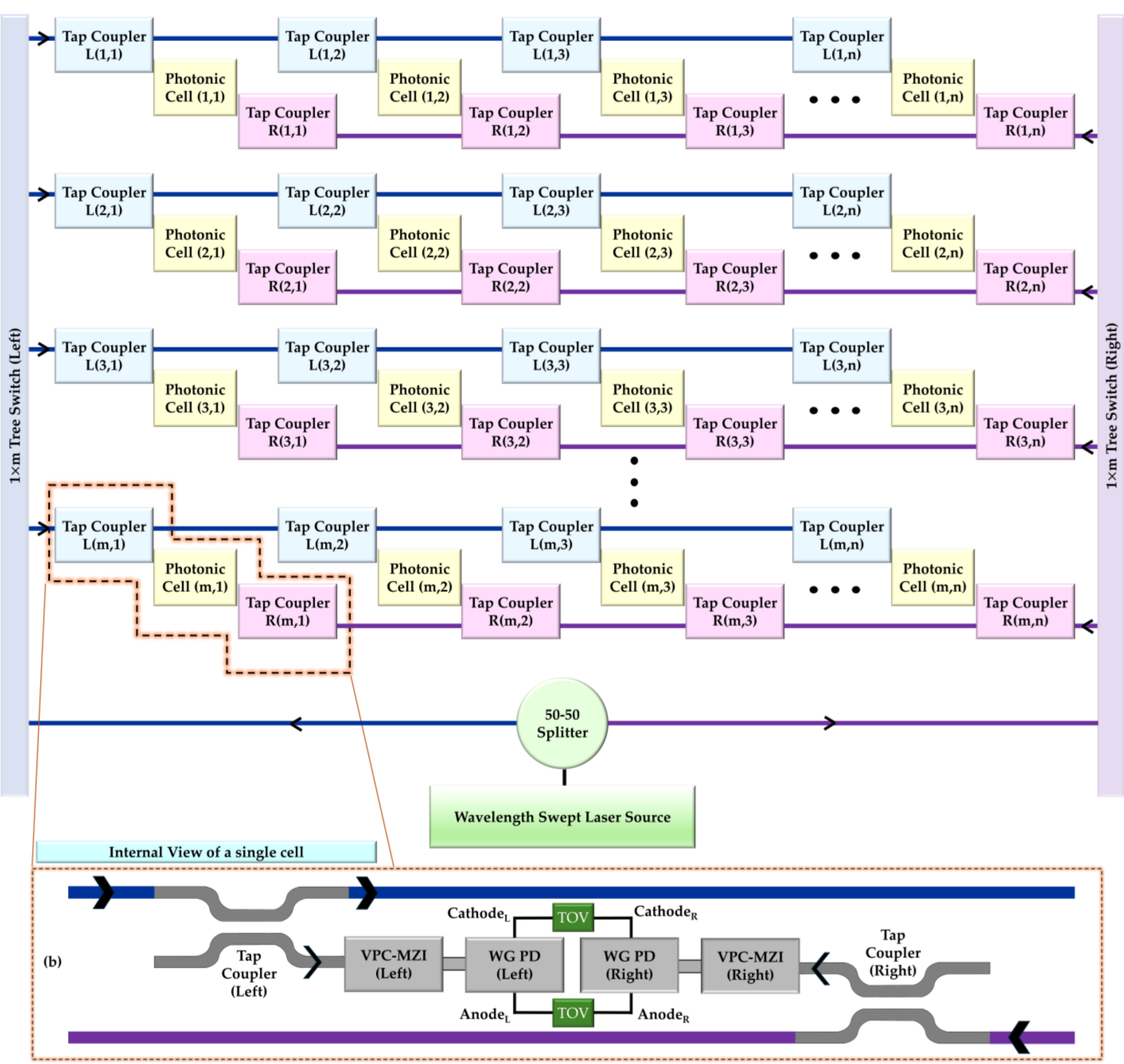

**Fig. 6: (a)** Photonic circuit layer architecture using identical tap ratios for all tap couplers. A wavelength-swept laser source feeds the circuit through a 50-50 splitter. The optical power propagates in both left and right directions along the waveguides. Bidirectional propagation enables a design where identical tap couplers can be used throughout the array. Each photonic cell receives light from two tap couplers, one from the left path and one from the right path. The leftmost photonic cells receive the highest power from the left tap couplers and the lowest power from the right tap couplers. The opposite occurs for the rightmost cells. **(b)** Internal view of a single photonic cell. Each cell contains two symmetric branches consisting of a tap coupler, a VPC-MZI sensor, and a WG photodetector. The optical signals from the left and right tap couplers are processed by the corresponding VPC-MZI and WG photodetector pairs. The photocurrents from the two photodetectors are summed electrically to generate the total photocurrent of the photonic cell. The photodetectors are connected to the CMOS layer through through-oxide-vias (TOVs) for electrical readout.

## VI. Alternative photonic ARRAY Architecture

An alternative photonic circuit layer architecture that uses identical tap couplers throughout the array is shown in Fig. 6(a). In contrast to the architecture in Section IV, where the tap ratios are intentionally varied along the row bus to equalize the delivered optical power, this implementation uses tap couplers with the same coupling ratio at every stage. The optical signal from the wavelength-swept laser source is first divided by a 50-50 splitter and injected into the row buses from both the left and right sides. The optical power then propagates bidirectionally along each row waveguide and is partially extracted by the tap couplers to feed the photonic cells. Because each row is illuminated from both directions, every photonic cell receives optical power from two independent propagation paths. As illustrated in Fig. 6(b), the internal structure of each photonic cell consists of two symmetric branches, each containing a tap coupler, a VPC-MZI sensor, and a WG photodetector. The left and right optical inputs generate photocurrents in their respective photodetectors, which are electrically combined to produce the total photocurrent of the pixel

$$I_{\text{total}} = \left(I_{\text{ph,1}} + I_{\text{dark,1}}\right) + \left(I_{\text{ph,2}} + I_{\text{dark,2}}\right)$$

This electrical summation allows the sensing signal to depend on the total optical power delivered from both directions. To illustrate the power distribution, consider a row containing n photonic cells. Let the optical power injected into the row from each side be $P_{in}/2$. If each tap coupler extracts a fraction $\kappa^2$ of the propagating optical power, then the optical power reaching the $j$-th cell from the left side can be approximated as

$$P_L(j) = \frac{P_{in}}{2}(1-\kappa^2)^{j-1}\kappa^2$$

The derivation of this power number is shown in supplementary material S2. Similarly, the optical power arriving from the right side is

$$P_R(j) = \frac{P_{in}}{2}(1-\kappa^2)^{(n-j)}\kappa^2$$

The total optical power received by the j-th photonic cell is therefore $P_{cell}(j) = P_L(j) + P_R(j)$. Because the two contributions originate from opposite directions, the variation of optical power along the row becomes significantly more uniform compared with a single-sided distribution scheme. Cells located near the left side receive more power from the left propagation path but less from the right path, while cells near the right side experience the opposite behavior. The middle cells receive comparable contributions from both directions. This bidirectional optical distribution also improves robustness against temperature-induced variations in passive optical components. Changes in temperature slightly modify the refractive index of the waveguide materials, which can affect the coupling ratios of directional couplers and splitters. In the proposed architecture, any deviation in coupling ratio that affects power delivery from one direction is partially compensated by the contribution from the opposite direction. Because the photocurrents from the two photodetectors are electrically summed within the pixel circuit, the effective sensing signal depends on the total optical power received from both sides. This combination of bidirectional illumination and electrical summation reduces sensitivity to coupling-ratio drift and helps maintain stable optical power delivery across the array.

Overall, the alternative architecture provides a simple implementation using identical tap couplers while improving tolerance to coupling variations and maintaining a more balanced optical power distribution across the photonic sensor array. However, this scheme requires duplicating the sensing elements within each cell (two VPC-MZI and two photodetectors) to support bidirectional illumination, resulting in higher component count compared to the architecture presented in Section IV. Consequently, the two architectures offer different trade-offs between implementation simplicity, robustness to coupling variations, and device count, and can be selected based on the requirements of the target application.

## VII. PDP, Pitch Estimation & Benchmarking

The sensing power-delay product (PDP) of the proposed ThermoPix circuit is evaluated from the transient simulation waveforms. The sensing delay is defined as the time difference between the 50% falling edge of the DEN signal and the 50% falling edge of the output signal OUT, which corresponds to the moment when the sensing circuit detects the threshold-crossing event. The average power consumption during this sensing transition is measured from the HSPICE simulations. The PDP is then calculated by multiplying the measured sensing delay with the corresponding power consumption. Using this methodology, the sensing PDP of the ThermoPix circuit is obtained as 0.152 fJ.

The photonic cell pitch of the proposed architecture depends on the arrangement of the photonic components within each cell. In the first approach, the tap coupler, VPC-MZI sensor, and WG photodetector are placed sequentially along the row waveguide. Based on the component dimensions presented in Section II-B and Section IV, the resulting average photonic cell pitch is 23.26 µm. The value is reported as an average because the coupling length of the directional tap couplers varies along the row in order to achieve the required tap ratios for uniform optical power distribution. In the second approach, the architecture employs bidirectional optical feeding, which requires two symmetric photonic branches within each cell. As a result, the number of photonic components per cell is doubled, leading to an increased pitch. In this implementation, identical tap couplers with a fixed minimum coupling length are used, and the resulting photonic cell pitch becomes 38.52 µm. These values reflect the trade-off between compact layout in the first architecture and improved robustness to coupling variations in the second architecture.

Table II compares the proposed ThermoPix architecture with representative temperature sensing approaches reported in the literature [2], [35], [36], [37], [38]. Most prior CMOS temperature sensors rely on analog or mixed-signal sensing

| TABLE II: Benchmarking | | | | | | | |
|---|---|---|---|---|---|---|---|
| **Reference** | **Sensor Type** | **System Style** | **Tech Node** | **Readout / Conversion Time** | **Sensitivity** | **Energy / Power Metric** | **Pixel Pitch** |
| [35] | Time-domain CMOS sensor | Electronic | 0.18 µm | 22.75 ms | N/A | FoM 6.7 pJ/K² | N/A |
| [2] | MOS temperature sensor | Electronic | 0.18 µm | 16 ms | ±0.75 °C accuracy | 36–40 µW | 11 µm |
| [36] | Diode temperature sensor | Electronic | 0.35 µm | ms-scale scan | ~1 mV/K | N/A | ~250 µm |
| [37] | CMOS thermal sensor array | Electronic | 0.18 µm | ms scale | N/A | N/A | N/A |
| [38] | Photonic ring resonator | Photonic | 0.18 µm | ~6 µs response | 83 pm/°C | N/A | N/A |
| **ThermoPix (Approach I)** | Photonic thermal sensor array | Electronic–photonic 3D | 14 nm | **2 µs row readout** | **3.15 ns/K** | **0.152 fJ PDP** | **23.26 µm** |
| **ThermoPix (Approach II)** | Photonic thermal sensor array | Electronic–photonic 3D | 14 nm | **2 µs row readout** | **3.15 ns/K** | **0.152 fJ PDP** | **38.52 µm** |

circuits, which are commonly implemented in older technology nodes such as 0.18 µm or 0.35 µm CMOS. In contrast, the ThermoPix readout circuit is largely digital, with the thresholding function assisted by the PTM device. This allows the design to leverage a scaled CMOS node (14 nm), enabling lower switching energy and contributing to the low sensing PDP reported for the circuit. Another key distinction lies in the array implementation. Several prior works demonstrate temperature sensing either as single-point sensors or as electronic sensor arrays, whereas ThermoPix performs temperature sensing through a photonic layer integrated with CMOS readout circuitry. In this architecture, the spatial pitch of the sensing elements is primarily determined by the photonic components rather than the CMOS circuitry. As a result, the achievable pixel pitch is governed by the dimensions of the photonic devices and routing structures, while the CMOS layer mainly provides compact timing-based readout and digital processing. These differences highlight the distinct design trade-offs between conventional electronic temperature sensors and the proposed electronic-photonic ThermoPix architecture.

## VIII. Conclusion

This paper presents ThermoPix, a temperature sensing architecture that integrates a photonic sensing layer with CMOS timing-based readout circuitry in a 3D electronic–photonic platform. We convert temperature-induced spectral shifts in a VPC-MZI sensor into timing information that can be efficiently processed by compact CMOS circuits. Through circuit-level simulations, we demonstrate microsecond-scale row readout with a sensing PDP of 0.152 fJ and a sensitivity of 3.15 ns/K, while requiring only 150 nW optical power per photonic cell. We also investigate two alternative photonic circuit layer architectures for optical power distribution, offering different trade-offs between compact pixel pitch and tolerance to coupling variations. Compared with prior approaches, we combine photonic sensing with CMOS readout in an array-based architecture. While conventional CMOS temperature sensors typically exhibit millisecond-scale conversion times and photonic sensors are often demonstrated as single-point devices, our approach enables array-level temperature sensing with microsecond-scale row readout.

In future work, we will further evaluate and extend the proposed architecture. We will perform process variation analysis to study the impact of variations in the photonic components, PTM device, and CMOS circuitry on sensing accuracy and timing stability. We will also explore circuit-level optimization to improve robustness and reduce energy consumption of the readout operation. In addition, we will investigate array scaling, improved photonic routing architectures, and experimental validation of the proposed concept using integrated electronic-photonic platforms. These directions may further establish ThermoPix as a scalable approach for high-resolution on-chip temperature monitoring in advanced electronic systems.